\documentclass[aps,prb,twocolumn,superscriptaddress,showpacs]{revtex4}
\usepackage{graphicx}
\usepackage{dcolumn}
\usepackage{bm}
\usepackage{amsmath}

\begin{document}


\title{
Multipole tensor analysis of the resonant x-ray scattering by quadrupolar and magnetic order in DyB$_2$C$_2$}


\author{T. Matsumura}
\email[]{tmatsu@iiyo.phys.tohoku.ac.jp}
\affiliation{Department of Physics, Graduate School of Science, Tohoku University, Sendai, 980-8578, Japan}
\author{D. Okuyama}
\affiliation{Department of Physics, Graduate School of Science, Tohoku University, Sendai, 980-8578, Japan}
\author{N. Oumi}
\affiliation{Department of Physics, Graduate School of Science, Tohoku University, Sendai, 980-8578, Japan}
\author{K. Hirota}
\affiliation{Institute for Solid State Physics, The University of Tokyo, Kashiwanoha, Kashiwa, 277-8581, Japan}
\author{H. Nakao}
\affiliation{Department of Physics, Graduate School of Science, Tohoku University, Sendai, 980-8578, Japan}
\author{Y. Murakami}
\affiliation{Department of Physics, Graduate School of Science, Tohoku University, Sendai, 980-8578, Japan}
\author{Y. Wakabayashi}
\affiliation{Institute of Materials Structure Science, High Energy Accelerator Research Organization, Tsukuba, 305-0801, Japan}


\date{\today}

\begin{abstract}
Resonant x-ray scattering (RXS) experiment has been performed for the (3 0 1.5) superlattice reflection in the 
antiferroquadrupolar and antiferromagnetic phase of DyB$_2$C$_2$. 
Azimuthal-angle dependence of the resonance enhanced intensities for both dipolar ($E1$) and quadrupolar ($E2$) resonant 
processes has been measured precisely with polarization analysis. Every scattering channel exhibits 
distinctive azimuthal dependence, differently from the symmetric reflection at (0 0 0.5) which was studied previously. 
We have analyzed the results using a theory developed by Lovesey \textit{et al.}, which directly connects atomic tensors 
with the cross-section of RXS. The fitting results indicate that the azimuthal dependences can be explained well by 
the atomic tensors up to rank 2. Rank 3 and rank 4 tensors are reflected in the data very little. In addition, 
The coupling scheme among the $4f$ quadrupolar moment, $5d$ ortitals, and the lattice has been determined 
from the interference among the Thomson scattering from the lattice distortion and the resonant scatterings of $E1$ and $E2$ 
processes. It has also been established from the RXS of the (3 0 1.5) reflection that 
the canting of the $4f$ quadrupolar moments exists up to $T_{Q}$. We also discuss a possible wavefunction of the 
ground state from the point-charge model calculation. 
\end{abstract}


\maketitle


\section{Introduction} 
Anisotropic charge distributions of localized $4f$ electrons play important roles in magnetic 
properties of rare-earth compounds through quadrupolar interactions and orderings since the magnetic 
moments are strongly coupled with the quadrupolar moments by the spin-orbit interaction. 
Unusual magnetic phase diagrams and field induced antiferromagnetic structures 
typically observed in CeB$_6$, TmTe, DyB$_2$C$_2$, etc., cannot be understood without considering 
the quadrupolar moments.~\cite{Effantin85,Link98,Mignot00,Mignot02,Yamauchi99} 
In CeB$_6$ even an octupolar moment is necessary to explain the 
experimental observations.~\cite{Shiina97,Shiina98,Kusunose01} 
What is more exotic, in NpO$_2$, it is claimed that the $\Gamma_5$-type octupolar order is realized 
without a magnetic dipolar moment.~\cite{Paixao02}
It is therefore very important to study various behavior of quadrupolar and  higher rank multipolar moments. 

\begin{figure}[tb]
\begin{center}
\includegraphics[width=7.5cm]{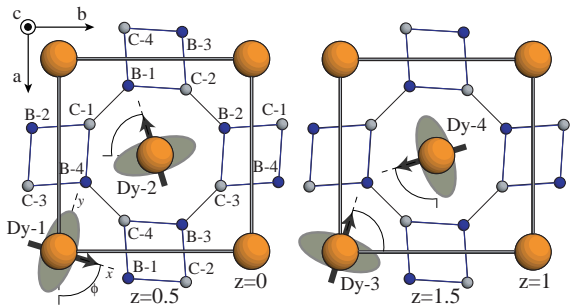}
\end{center}
\caption{Top view of the crystal structure of DyB$_2$C$_2$ ($P4/mbm$, $a=5.341$ \AA, $c=3.547$ \AA\ at 30 K). 
Arrows indicate the magnetic moments in the AFM phase. Ellipses indicate the schematic views of 
pancake-like charge distributions of Dy below $T_{\text Q}$. Their flat planes are perpendicular to the magnetic moments. 
The B-C layers are shifted from the Dy layers by $z=0.5$. 
}
\label{fig1}
\end{figure}

Recently, a series of tetragonal rare-earth compound of RB$_2$C$_2$ (R=Tb, Dy, Ho) has been studied extensively as 
a typical system which exhibits antiferroquadrupolar (AFQ) orderings and complex magnetic 
structures.~\cite{Yamauchi99,Ohoyama00,Kaneko01,Kaneko02, Ohoyama01} 
Among the three compounds DyB$_2$C$_2$ is the most typical AFQ system with the highest $T_{Q}$ at 24.7 K 
and a well separated $T_{N}$ at 15.3 K, indicating that the quadrupolar interaction is stronger than the magnetic interaction. 
In HoB$_2$C$_2$ and TbB$_2$C$_2$ the two interactions are comparable in magnitude and compete 
with each other, leading to complex magnetic structures and phase diagrams.~\cite{Ohoyama01} 
The magnetic structure of DyB$_2$C$_2$ in the antiferromagnetic (AFM) phase below $T_{N}$ 
determined by Yamauchi \textit{et al.} is shown in Fig.~\ref{fig1}.~\cite{Yamauchi99} The moments along the $c$ axis are 
almost perpendicular with each other and those in the $ab$ plane are antiparallel but canted from the $\langle$1 1 0$\rangle$ 
directions. This structure is characterized by four $\bm{k}$-vectors: $\bm{k}_1$=(1 0 0), 
$\bm{k}_2$=(1 0 0.5), $\bm{k}_3$=(0 0 0), and $\bm{k}_4$=(0 0 0.5). This unusual magnetic structure 
can be understood naturally by considering the underlying AFQ order. 

Observation of the AFQ order in DyB$_2$C$_2$ has been performed to date by resonant x-ray scattering (RXS). 
By tuning the energy of the incident photon to an absorption edge of the element in study, a core electron 
is promoted to an unoccupied valence shell, forming an intermediate state. Secondary photon is 
emitted when the electron returns to the core state, and the scattered photon contains information of the 
valence shell. At the $L_{\text{III}}$ 
absorption edge of the rare-earth elements the intermediate state is described by a $2p_{3/2}$ core hole 
and an additional electron in the $5d$ shell ($E1$ process) or $4f$ shell ($E2$ process). 

After the first observations of RXS in DyB$_2$C$_2$ by Hirota \textit{et al.} and Tanaka \textit{et al.},~\cite{Hirota00,Tanaka99} 
we made a detailed analysis on the observed reflections using a theory developed by Blume.~\cite{Matsumura02,Blume94} 
The experimental results are well explained by the model of the AFQ structure 
shown in Fig.~\ref{fig1}. In Blume's theory, the scattering-factor tensor is devided into symmetric and 
asymmetric parts, which are interpreted in Ref.~\onlinecite{Matsumura02} as quadrupolar and magnetic origin, respectively. 
The scattering amplitudes calculated for each $\bm{k}$-vector, polarization channel, and 
transition process ($E1$ and $E2$), are mostly consistent with the observed results. 

In Ref.~\onlinecite{Matsumura02}, we suggested that the canting angle of the 
quadrupolar moment above $T_{N}$ could be zero; i.e., they align along the [1 1 0] directions. 
This statement came from the experimental result that the (1 0 2.5) resonant reflection for the 
$\sigma$-$\pi'$ channel seemed to disappear above $T_{N}$. The calculated 
scattering amplitude of this reflection in the AFQ phase contains the symmetric part of the scattering-factor 
tensor multiplied by $\sin 2\alpha$, where $\alpha$ is the canting angle from the [1 1 0] directions.  
Since the former factor is definitely not zero in the AFQ phase, the latter factor $\alpha$ must be zero 
if the experimental result is correct. This is an important problem when we consider the origin of the 
canting in detail; one is an inter-ionic quadrupolar interaction mediated by the conduction electrons 
and the other is a quadrupole-strain interaction by the tilts and shifts of the surrounding B-C octagon. 
Note that the B-C octagons around Dy-1 and Dy-2 are oppositely tilted about the $c$ axis. 
If $\alpha=0^{\circ}$ above $T_{N}$, we would come to a statement that the inter-ionic quadrupolar interaction 
is much stronger than the quadrupole-strain interaction and prefer $90^{\circ}$ arrangement along the $c$ axis 
and parallel arrangement in the $ab$ plane.

Lattice distortion below $T_{Q}$ characterized by the propagation vector of (1 0 0.5) is also important in 
DyB$_2$C$_2$. Although this lattice distortion should be compatible with the AFQ order, 
we tried to discuss a possibility of the motion of Dy atoms in the previous papers.~\cite{Hirota00,Matsumura02} 
We compared the structure factors calculated for two models of the subgroup of $P4/mbm$; in one model the Dy 
atoms are shifted ($P4/mnc$, No. 128) and in the other the B-C atoms are shifted ($P4_2/mnm$, No. 136), 
both along the $c$ axis. 
In $P4_2/mnm$ B-1 and B-3 in Fig.~\ref{fig1} shift upward and B-2,4 shift downward. 
Motions of C atoms have two possibilities: C-1,3 up and C-2,4 down or C-2,4 up and C-1,3 down. 
We calculated the former case in Ref.~\onlinecite{Matsumura02} and claimed that the observed intensity of 
the (1 0 2.5) reflection cannot be obtained. However, the latter case explains the observed intensity 
by assuming the shift of about $10^{-3} c$. 
Later on, Adachi \textit{et al.} performed nonresonant x-ray diffraction experiment and observed (0 1 half-integer), 
(1 1 half-integer), and (1 2 half-integer) reflections. The intensities are well explained by assuming the 
B-C motion of the latter case.~\cite{Adachi02} The motion of Dy in $P4/mnc$ forbids the (1 1 half-integer) reflections, 
which contradicts the experimental result. 
It should be noted that the validity of the analysis of the RXS data is not affected because 
it assumes the site symmetry of $2/m$ which is compatible with the space group $P4_2/mnm$.

In the previous works of RXS in DyB$_2$C$_2$, the (0 0 half-integer) reflections for the $\sigma$-$\sigma'$ 
channel in the $E2$ process were hidden in the tail of the $E1$ resonance, while those for the $\sigma$-$\pi'$ 
channel were well resolved.~\cite{Hirota00,Tanaka99,Matsumura02} 
Recently, Tanaka \textit{et al.} made a systematic RXS study of the 
(0 0 half-integer) reflections and analyzed the data using the structure factor of atomic tensors 
calculated by Lovesey \textit{et al.}~\cite{Tanaka04,Lovesey01} 
They ascribed the disappearance of the $E2$ resonance in the 
$\sigma$-$\sigma'$ channel to the idea that the rank 2 (quadrupole) and rank 4 (hexadecapole) tensors 
cancel out in the structure factor. Although they claim this as the observation of the rank 4 tensor,  
it should be checked by more detailed experiments because such a high rank multipole 
should be difficult to observe experimentally. 
We have shown in another paper that the rank 4 multipole is not necessary 
if we consider the interference between $E1$ and $E2$ resonances.~\cite{Matsumura04} 
Further evidence that support this statement will be presented in this paper.

In the present study, we have performed RXS experiment on the (3 0 1.5) reflection 
to investigate the above problems; one is the canting angle in the AFQ phase and the other is the 
possibility of observing the higher rank atomic tensors.  
We have analyzed the data using a theory developed by Lovesey \textit{et al.}, 
which directly connects atomic tensors of different ranks with the cross-section of 
RXS.~\cite{Lovesey96a, Lovesey96b, Lovesey97, Lovesey98} 
A theory by Blume,~\cite{Blume94} or Hill and McMorrow,~\cite{Hill96} basically gives the same 
result as the one by Lovesey \textit{et al.} 
However, physical meanings of the parameters that appear in the formulae are not as clear as those 
in Lovesey's formulae.

\section{Experiment}
\begin{figure}[tb]
\begin{center}
\includegraphics[width=7.5cm]{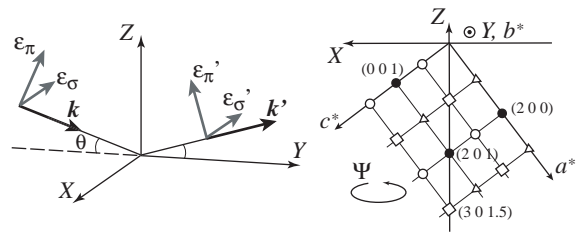}
\end{center}
\caption{Left: Configuration of the incident and scattered x rays in the laboratory coordinates $XYZ$. 
Right: Configuration of the reciprocal lattice space in the laboratory coordinates at the azimuthal angle $\Psi=0^{\circ}$. 
The sample rotation is counter-clockwise about $Z$ with increasing $\Psi$. 
}
\label{fig2}
\end{figure}
A single crystal of DyB$_2$C$_2$ was grown by the Czochralski method with a tetra-arc furnace. 
A flat (2 0 1) surface with its area about $1\times 2$ mm$^2$ was obtained by rubbing the crystal 
with sandpaper and then polishing the surface with fine emery paper. 
RXS experiments were performed on a four-circle diffractometer at the beamline 16A2 of the Photon Factory, KEK. 
The incident energy was tuned near the Dy $L_{\text{III}}$ edge. 
The sample was mounted in a closed-cycle helium refrigerator so that the (2 0 1) plane was normal to the $\phi$ axis 
of the diffractometer. The configuration of the sample and x rays are illustrated in Fig.~\ref{fig2}. 
The azimuthal-angle scan was performed by rotating the $\phi$ axis of the diffractometer. 
The mosaic width of the (2 0 1) fundamental reflection was about $0.06^{\circ}$. 

The incident x ray is almost linearly polarized with its electric field perpendicular to the scattering plane 
($\sigma$ polarization). To separate the $\sigma'$ (unrotated) and $\pi'$ ($90^{\circ}$ rotated) components 
of the diffracted beam, PG (0 0 6) reflection was used as an analyzer, for which the scattering angle becomes 
$90.7^{\circ}$ and almost perfect analysis ($\cos^2 90.7^{\circ}=1.5\times 10^{-4}$) is achieved. 
However, the contaminations caused by the unpolarized component of the incident beam cannot be neglected. 
The total contamination, estimated from the intensity ratio of $I_{\text{rotated}}/I_{\text{unrotated}}$ of the fundamental (2 0 1) 
reflection, was about 0.3\%.

\section{Results and analysis}
\subsection{Energy dependence}
\begin{figure}[t]
\begin{center}
\includegraphics[width=8cm]{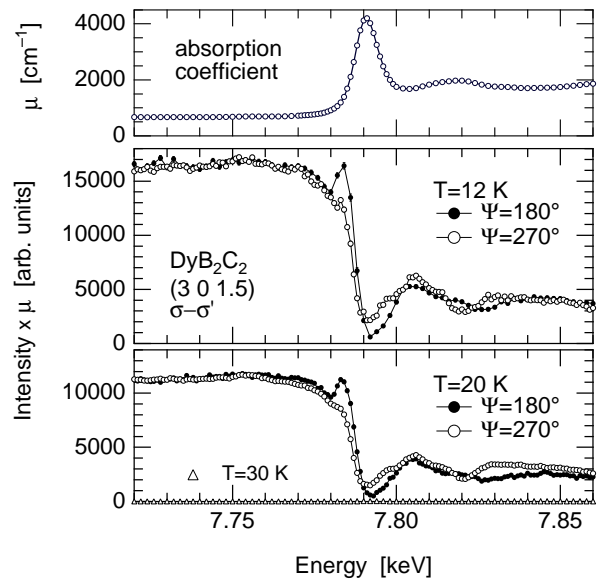}
\end{center}
\caption{Incident energy dependences of the intensity of the (3 0 1.5) reflection for the $\sigma$-$\sigma'$ channel 
at $\Psi=180^{\circ}$ and $270^{\circ}$. The energy dependence of the absorption coefficient 
determined from the fluorescence spectrum is shown in the top figure. 
}
\label{fig3}
\end{figure}
\begin{figure}[t]
\begin{center}
\includegraphics[width=8cm]{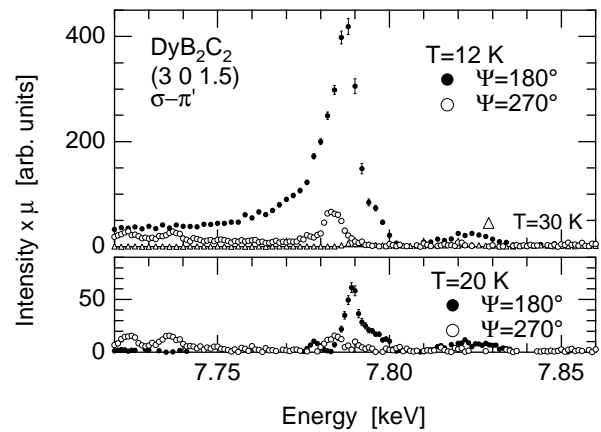}
\end{center}
\caption{Incident energy dependences of the intensity of the (3 0 1.5) reflection for the $\sigma$-$\pi'$ channel 
at $\Psi=180^{\circ}$ and $270^{\circ}$.  
}
\label{fig4}
\end{figure}
Figure \ref{fig3} shows the energy dependences of the (3 0 1.5) reflection at $T=12$ K in the AFM phase and at 
$T=20$ K in the AFQ phase. The data have been corrected for the absorption coefficient that was determined from the 
fluorescence spectrum.~\cite{Dumesnil98} The disagreement of the intensities at energies well below and above the edge 
is due to the imperfect correction. 
Large nonresonant contribution of the Thomson scattering from the lattice distortion is observed. 
The data are normalized to the intensity at 7.72 keV and $\Psi=180^{\circ}$ because this Thomson scattering does not 
exhibit azimuthal-angle dependence, which has been checked by comparing with the azimuthal scan of the 
(2 0 1) fundamental reflection. 

An azimuthal-angle dependence is clearly observed around the edge. 
It is remarkable that a large resonant enhancement is observed at 7.782 keV, which is ascribed to the $E2$ resonance. 
The enhancement of the $E1$ resonance at 7.792 keV is not clearly seen because the energy dependence of the 
nonresonant component due to the absorption is much stronger than that of the $E1$ resonance. However, the observed azimuthal 
dependence around 7.792 keV may be ascribed to the resonance of the $E1$ process. 

Figure \ref{fig4} shows the energy dependences for the $\sigma$-$\pi'$ channel. 
Contamination of the unrotated scattering has been subtracted and the data are normalized to the nonresonant 
intensity at 7.72 keV and $\Psi=180^{\circ}$ for the $\sigma$-$\sigma'$ channel. 
At 20 K, only the $E1$ resonance is observed at $\Psi=180^{\circ}$ and no signal is observed at $\Psi=270^{\circ}$. 
At 10 K, the spectrum at $\Psi=180^{\circ}$ exhibits large enhancement with a long tail to the lower energy side 
because of the interference between the nonresonant magnetic scattering and the resonant scatterings of 
$E1$ and $E2$ processes.~\cite{Gibbs91} The spectrum at $\Psi=270^{\circ}$ exhibits only the $E2$ resonance. 
In both $\sigma$-$\sigma'$ and $\sigma$-$\pi'$ channels, the resonant features clearly exist 
still at 20 K in the AFQ phase, which indicate that the the quadrupolar moment is still canted. The result is in accord with the 
site symmetry of $2/m$ and a symmetry analysis recently performed by Zaharko \textit{et al}.~\cite{Zaharko04}

\subsection{Azimuthal-angle dependence}
\begin{figure}[tb]
\begin{center}
\includegraphics[width=8cm]{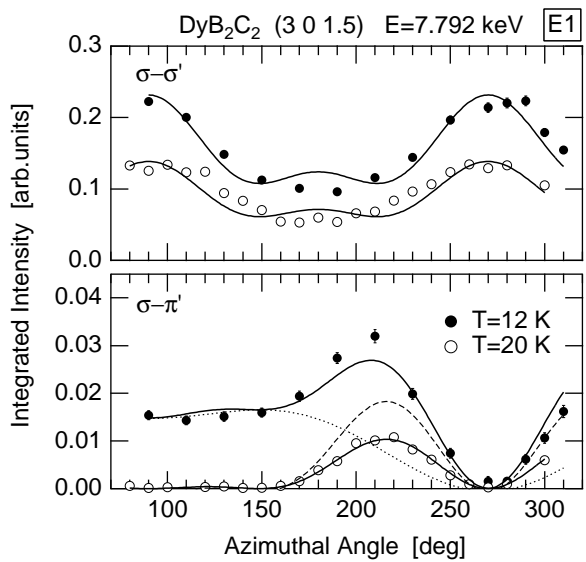}
\end{center}
\caption{Azimuthal-angle dependences of the integrated intensity of the (3 0 1.5) reflection for the $E1$ process. 
Solid lines are the fits described in the text. Dotted and dashed lines in the $\sigma$-$\pi'$ channel 
indicate the rank 1 and rank 2 contributions, respectively, to the fitting result at 12 K. 
}
\label{fig5}
\end{figure}
\begin{figure}[tb]
\begin{center}
\includegraphics[width=8cm]{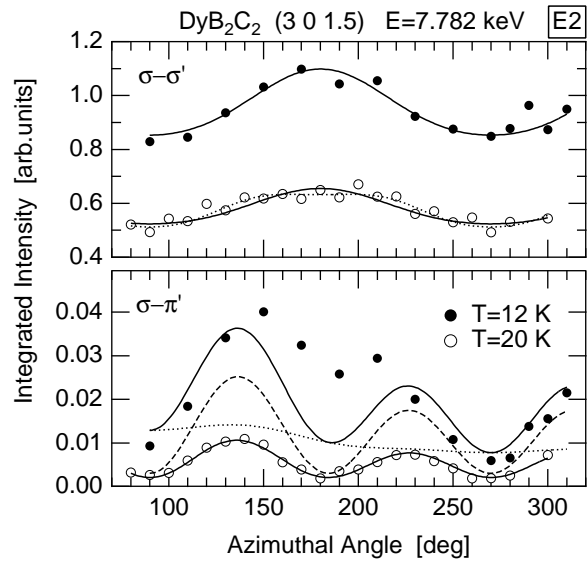}
\end{center}
\caption{Azimuthal-angle dependences of the integrated intensity of the (3 0 1.5) reflection for the $E2$ process. 
Solid lines are the fits described in the text. Dotted and dashed lines in the $\sigma$-$\pi'$ channel 
indicate the rank 1 plus rank 3 (magnetic) and rank 2 (quadrupolar) contributions, resspectively, 
to the fitting result at 12 K. The dotted line in the $\sigma$-$\sigma'$ channel indicate a fit including the 
rank 4 contribution. 
}
\label{fig6}
\end{figure}
Figure \ref{fig5} shows the azimuthal-angle dependence of the integrated intensity for the $E1$ process measured 
at 7.792 keV. The data are normalized to the nonresonant intensity at 7.72 keV for $\sigma$-$\sigma'$. 
In spite of the experimental condition that 7.792 keV corresponds to the minimum position of the 
energy spectrum in Fig.~\ref{fig3} because of the absorption, the intensity of the $\sigma$-$\sigma'$ channel 
varies with the azimuthal angle with $180^{\circ}$ periodicity. 
It is noted that the amplitude of the azimuthal dependence for the $\sigma$-$\sigma'$ channel is much 
larger than that for the $\sigma$-$\pi'$ channel. This is due to the interference with the strong Thomson scattering 
in the $\sigma$-$\sigma'$ channel, which will be explained in the next subsection. 

Figure \ref{fig6} shows the azimuthal-angle dependence for the $E2$ process measured at 7.782 keV. 
The data are again normalized to the nonresonant intensity at 7.72 keV for $\sigma$-$\sigma'$. 
Apparently, the azimuthal dependence is completely different from that of the $E1$ process. 
This fact also supports the ascription of this resonance to the $E2$ process, which measures the $4f$ state directly.  
This is different from the case in YTiO$_3$ where the azimuthal-angle dependences of the RXS intensity 
at the pre-edge and the main-edge are the same and the pre-edge resonance is ascribed to the 
$E1$ process.~\cite{Nakao02}

\subsection{Analysis}
\subsubsection{Azimuthal-angle dependence}
\begin{figure}[tb]
\begin{center}
\includegraphics[width=8cm]{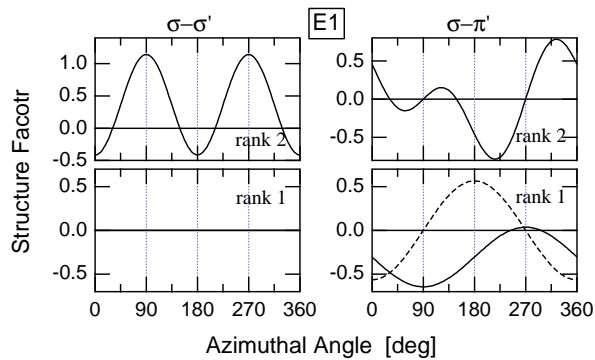}
\end{center}
\caption{Calculated azimuthal-angle dependences of the tensor structure-factors of the (3 0 1.5) reflection for the 
$E1$ resonance. The dashed line for the rank 1 tensor indicates the structure factor of the other domain. 
The domain only reverses the sign for the rank 2 tensor.  
The squares of these factors are compared with the data in Fig.~\ref{fig5}. 
}
\label{fig7}
\end{figure}
\begin{figure}[tb]
\begin{center}
\includegraphics[width=8cm]{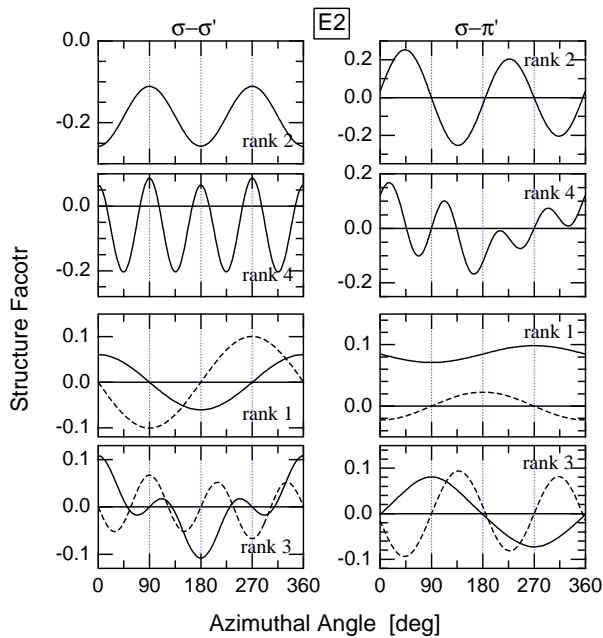}
\end{center}
\caption{Calculated azimuthal-angle dependences of the tensor structure-factors of the (3 0 1.5) reflection for the 
$E2$ resonance. The dashed lines for the odd rank  tensors indicate the structure factors of the other domain. 
The domain only reverses the sign for the even rank tensors.  
The squares of these factors are compared with the data in Fig.~\ref{fig6}. 
}
\label{fig8}
\end{figure}
In order to analyze the energy and azimuthal-angle dependences, we have calculated the structure factors 
of atomic tensors following a theory developed by Lovesey \textit{et al.}~\cite{Lovesey96a, Lovesey96b, Lovesey97, Lovesey98} 
The structure shown in Fig.~\ref{fig1} is assumed. The canting angle $\phi$ of the magnetic moments, 
which is assumed to coincide with the principal axis of the quadrupolar moment in the AFQ phase, 
is fixed at $73^{\circ}$ as determined in Ref.~\onlinecite{Yamauchi99}. Note that $\phi$ is an angle from 
the [1 0 0] axis in this paper. 
The spherical tensor $\langle T^{(K)}_q \rangle$ is defined in the local ionic coordinates $xyz$ as indicated in 
Fig.~\ref{fig1}. The direction of the magnetic moment is taken as the local $x$ axis. 

Since the Dy ion is in the local symmetry of $2/m$ in the AFQ phase, $\langle T^{(K)}_q \rangle$ 
with $K=$even has only the components of $q=\pm 2$ and the relation 
$\langle T^{(K)}_2 \rangle=\langle T^{(K)}_{-2} \rangle$ is satisfied. Both $\langle T^{(2)}_2 \rangle$ and 
$\langle T^{(4)}_2 \rangle$ are real, and $\langle T^{(2)}_2 \rangle$ represents the $\frac{\sqrt{3}}{2}(x^2-y^2)$ type 
quadrupolar moment and $\langle T^{(4)}_2 \rangle$ represents the 
$\frac{7\sqrt{5}}{4}\{x^4-y^4-\frac{6}{7}(x^2-y^2)r^2\}$ type hexadecapolar moment. 
With respect to the odd rank tensors, since we assume the dipolar moments lie in the $ab$ plane, 
only $\langle T^{(K)}_q \rangle$ with $q=$odd is allowed and the relation 
$\langle T^{(K)}_q \rangle=-\langle T^{(K)}_{-q} \rangle$ is satisfied. Then, for $K=3$, only 
$\frac{5}{2}(x^3 - \frac{3}{5}xr^2)\equiv T_{1u}^{x}$ type and $\frac{\sqrt{15}}{2}x(y^2-z^2)\equiv T_{2u}^{x}$ 
type octupolar moments are allowed. 

The structure factors of the atomic tensors for the $E1$ and $E2$ processes are obtained by summing 
$Z^{(E1)}$ and $Z^{(E2)}$ for four Dy ions in the unit cell together with the phase factors as described in the 
Appendix. 
The azimuthal-angle dependences of the calculated structure factors $F^{(E1)}$ and $F^{(E2)}$ are 
shown in Figs.~\ref{fig7} and \ref{fig8}, respectively. The constant factors $A$ and $B$ are not considered. 
It should be noted that $\langle T^{(K)}_q \rangle$ in $F^{(E1)}$ and $F^{(E2)}$ represent 
atomic tensors of the $5d$ and $4f$ shells, respectively, and it is written as $\langle T^{(K)}_q \rangle_{5d}$ 
or $\langle T^{(K)}_q \rangle_{4f}$ when we specify the shell.

For even rank tensors with $K=$2 and 4, the coefficients of $\langle T^{(2)}_2 \rangle$ and 
$\langle T^{(4)}_2 \rangle$, respectively, are shown. For odd rank tensors with $K=$1 and 3, the 
coefficients of $i\langle J_x \rangle$ and $i\langle T_{1u}^{x} \rangle$, respectively, are shown. 
The squares of these structure factors are compared with the experimental results. 
The fitting results are shown by the solid lines in Figs.~\ref{fig5} and \ref{fig6}. 
At 20 K in the AFQ phase only the even rank tensors contribute to the scattering and at 10 K in the 
AFM phase the odd rank tensors of magnetic origin are superimposed. Since the odd rank tensors exhibit 
the domain structure, the intensities from the two domains are averaged in the fitting procedure. 

With respect to the $E1$ process, the data at 20 K are fitted with the structure factor of rank 2 only. 
The data for both polarization channels are well reproduced by the calculation, 
although there is a slight difference in the $\sigma$-$\sigma'$ channel around $\Psi=180^{\circ}$. 
At 12 K, the structure factor of rank 1 is added in the fitting, and the data for $\sigma$-$\pi'$ channel are 
well reproduced. The data for $\sigma$-$\sigma'$ channel still exhibit the azimuthal dependence of 
the rank 2 tensor because there is no contribution from the rank 1 tensor. 

With respect to the $E2$ process, the data at 20 K are well reproduced by the structure factor of rank 2 only. 
Although the rank 4 tensor should in principle contribute to the scattering, the data shows only the feature 
of the rank 2 tensor. The present scattering vector of (3 0 1.5) has different azimuthal-angle dependences for 
rank 2 and rank 4 tensors. Then, the introduction of the rank 4 tensor only disturbes the 
fitting results in Fig.~\ref{fig6}.  For $\sigma$-$\sigma'$ the fitting seems to be slightly 
improved by introducing the rank 4 tensor about 15\% as large as the rank 2 tensor. The result 
is shown by the dotted line, but this is within the experimental error. 
It is remarkable that the data for $\sigma$-$\pi'$ at 20 K are almost perfectly reproduced 
only by the rank 2 tensor. 

The data at 12 K for the $E2$ process involve even and odd rank tensors. However, for the $\sigma$-$\sigma'$ channel, 
the data were fitted only by the rank 2 tensor because it was not possible to discriminate the azimuthal 
dependence of the odd rank tensors. The data for the $\sigma$-$\pi'$ channel were fitted roughly by the rank 2, rank 1 
and rank 3 tensors. However, the intensity at 7.782 keV for the $\sigma$-$\pi'$ channel involves relatively large 
contribution from the interference between the nonresonant magnetic scattering and the $E1$ resonance. Then, 
the data strongly reflects the azimuthal dependence of the $E1$ resonance, 
and it is difficult to discriminate the pure $E2$ resonance.

\subsubsection{Interference effect}
\begin{figure}[tb]
\begin{center}
\includegraphics[width=8cm]{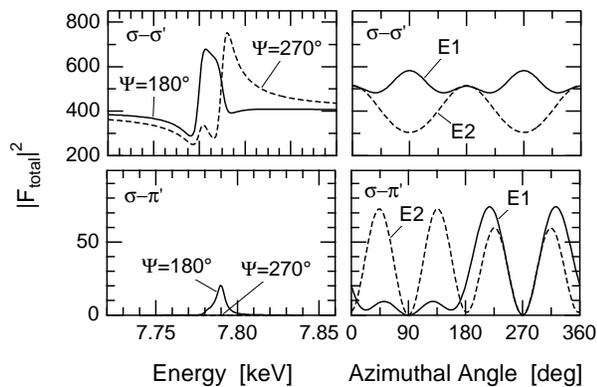}
\end{center}
\caption{Simulation of the energy and azimuthal-angle dependences of the (3 0 1.5) reflection 
near the absorption edge, considering the interference among the Thomson scattering and the resonances of $E1$ and 
$E2$ processes. The lattice does not contribute to the $\sigma$-$\pi'$ channel. The parameters 
are taken to be $F_{\text{lat}}=20$, $F^{(E2)}=F^{(E2)}_{\text{calc}}\times(-0.1)$,  
$F^{(E1)}=F^{(E1)}_{\text{calc}}\times 0.033$, $\Gamma_{E1}=\Gamma_{E2}=6$ eV, 
$\Delta_{E2}=7.78$ keV, and $\Delta_{E1}=7.79$ keV. 
}
\label{fig9}
\end{figure}
We have so far analyzed the azimuthal-angle dependences of the $E1$ and $E2$ processes independently. 
With respect to the data at 20 K in the AFQ phase, only the rank 2 tensor is enough to explain 
the data. However, we should be careful about the interference between $E1$ and $E2$ resonances. 
It is not obvious if the intensity at 7.782 and 7.792 keV reflects only the $E2$ and $E1$ resonance, 
respectively. In the (0 0 half-integer) reflection the interference between $E1$ and $E2$ resonances 
affects the energy spectra for $\sigma$-$\sigma'$ and $\sigma$-$\pi'$ channels as we have 
shown in a former paper.~\cite{Matsumura04}  
Furthermore, in the $\sigma$-$\sigma'$ channel the nonresonant Thomson scattering from the 
lattice distortion also interfares. We have to consider 
\begin{equation}
\left| F_{\text{lat}}+\frac{r_{2}F^{(E2)}}{E-\Delta_{E2}+i\Gamma_{E2}/2}
+\frac{r_{1}F^{(E1)}}{E-\Delta_{E1}+i\Gamma_{E1}/2} \right|^2
\label{eq:intfare}
\end{equation} 
for the $\sigma$-$\sigma'$ channel. $F_{\text{lat}}$ is not necessary for the $\sigma$-$\pi'$ channel. 
$r_{1}$ and $r_{2}$ are the mixing parameters to model more realistic oscillators of $E1$ and $E2$ resonances. 
In the present analysis, however, they are both fixed at unity. This point will be discussed in the next section. 

Figure \ref{fig9} shows a simulation of the energy and azimuthal-angle dependences 
in which the interference effect is considered. The calculated structure factors in Figs.~\ref{fig7} and \ref{fig8} are used. 
The features at 20 K in Figs.~\ref{fig3}, \ref{fig4},  \ref{fig5}, and  \ref{fig6} are well explained, 
although the $E1$ resonance Fig.~\ref{fig3} at $\Psi=270^{\circ}$ is hidden in the strong dip due to the absorption effect. 
With respect to $\sigma$-$\sigma'$, the scale of the longitudinal axis is much larger than that of $\sigma$-$\pi'$ because 
the Thomson scattering from the lattice distortion interferes. 

In order to reproduce the experimental results, the signs of $\langle T^{(2)}_2 \rangle_{4f}$ and 
$\langle T^{(2)}_2 \rangle_{5d}$ must be opposite and the signs of $\langle T^{(2)}_2 \rangle_{4f}$ 
and $F_{\text{lat}}$ must also be opposite. 
In Fig.~\ref{fig9} we assumed $\langle T^{(2)}_2 \rangle_{4f}=-0.1$ and $\langle T^{(2)}_2 \rangle_{5d}=0.033$. 
The opposite sign between the two tensors is consistent with the result of the RXS of the (0 0 2.5) reflection in Ref.~\onlinecite{Matsumura04} 
and the relation with $F_{\text{lat}}$ is newly obtained in the present reflection of (3 0 1.5). 
We note that the values of $\langle T^{(2)}_2 \rangle_{4f}$ and $\langle T^{(2)}_2 \rangle_{5d}$ used here are slightly different from 
those obtained in Ref.~\onlinecite{Matsumura04} for the (0 0 2.5) reflection. This is because the energy dependence of the atomic 
scattering factor is not common, but depends on the reflection points, as observed, e.g., in YTiO$_3$.~\cite{Nakao02}

The matrix element of $O^{(2)}_{2}=\frac{\sqrt{3}}{2}(J_x^{\;2}-J_y^{\;2})$ in the local coordinate becomes positive 
for the charge distribution of the $4f$ electrons shown in Fig.~\ref{fig1}, which is extended along the $y$ direction.  
However, since the reduced matrix element (r.m.e) of $T^{(2)}$ for the $4f$ electrons of Dy$^{3+}$ is negative, 
$\langle T^{(2)}_2 \rangle_{4f}$ becomes negative. Concerning the $5d$ orbital, we should 
consider $yz$ and $zx$ orbitals which are degenerate at $T>T_Q$. Below $T_Q$ the degeneracy is 
lifted by the quadrupolar moment of the $4f$ orbital through the $d$-$f$ Coulomb interaction 
or by the lattice distortion of B and C atoms through the crystal field effect. 
Both have the local symmetry of $2/m$, leading to the appearance of $\langle T^{(2)}_2 \rangle_{5d}$ below $T_Q$. 
The above analysis indicates that the sign of $\langle T^{(2)}_2 \rangle_{5d}$ is positive. 
Calculation of $\langle T^{(2)}_2 \rangle_{5d}$ using a formula in Ref.~\onlinecite{Lovesey97} 
shows that the $zx$ orbital gives positive and $yz$ orbital gives negative values with the same 
magnitude. This indicates that the $zx$ component in the occupied state increases in the AFQ phase. 
Therefore, the charge distributions of $4f$ and $5d$ orbitals are orthogonal with each other. 

Finally, the requirement that the sign of $F_{\text{lat}}$ must be positive indicates that the B-C parallelogram of 
B-2, B-4, C-1, and C-3, is shifted up and that of B-1, B-3, C-2, and C-4, is shifted down in the layer of 
$z=0.5$; the direction of the shifts is opposite in the layer of $z=1.5$. 
In the opposite case the sign of $F_{\text{lat}}$ becomes negative, which is inconsistent with the
experimental result. Then, the B-C atoms along the direction where the $4f$ charge distribution 
is extended move away from the Dy ion, while those along the direction where the 
$5d$ charge distribution is extended move toward the Dy ion.

\section{Discussions}
\subsection{multipolar moments}
\begin{figure}[t]
\begin{center}
\includegraphics[width=8cm]{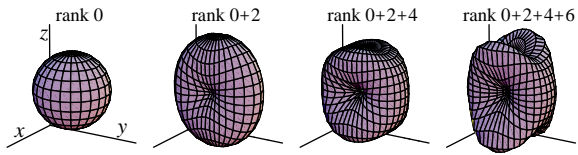}
\end{center}
\caption{Calculated charge distribution of the $4f$ electrons of a Dy ion in the AFQ phase. 
The figure is shown in the ionic coordinates and the $x$-axis corresponds to the direction of the 
magnetic moment in the AFM phase. 
Asphericity is emphasized by subtracting the spherical charge distribution of a Gd ion with seven $4f$ electrons. 
}
\label{fig10}
\end{figure}

The experimental results exhibit clear evidence of the rank 2 tensor, 
but the rank 4 tensor is little reflected in the data. 
Let us discuss the $4f$ wavefunction in the AFQ phase and the multipolar moments. 
In order to simulate the AFQ phase, we calculated a set of eigenfunctions of a crystal field 
Hamiltonian by the point charge model. Since the entropy at $T_Q$ is $R \ln 4$,~\cite{Yamauchi99} the ground state must be 
a pseudo-quartet where the first excited doublet has much smaller energy than the second one. 
In addition, since the dipolar moment has large components within the $ab$ plane, the main 
components of the eigenfunctions of the pseudo-quartet must have small $J_z$.
The eigenfunctions which satisfy these conditions were obtained by appropriately selecting the effective 
point charges.  Next, in the local coordinates, an $O^{(2)}_{2}$-type perturbation was added to make 
the eigenstates in the local symmetry of $2/m$.
Linear combinations within the pseudo-quartet are made and the eigenfunctions 
which diagonalize $O^{(2)}_{2}$ were obtained. 
The angular part of the charge distribution calculated for thus obtained ground doublet is demonstrated in 
Fig.~\ref{fig10}. The charge distribution is expressed by the even rank spherical harmonics up to rank 6, 
and the figure is shown with increasing the rank step by step. This eigenfunction has a magnetic dipolar moment 
of $9.5\mu_{B}$ along the $x$ direction, which is consistent with the experimental result.~\cite{Indoh00} 

The higher rank multipolar moments $O^{(K)}_{2}$ with even $K$ are simultaneously diagonalized. 
The expectation values of the operator equivalent $O^{(K)}_{2}$ for $K=2$ and 4 are listed in 
Table~\ref{tabl1} together with the r.m.e for the atomic tensor $T^{(K)}$ that appear in the scattering factor of the 
$E2$ resonance.~\cite{Lovesey98} The expectation value of $\langle T^{(K)}_{2} \rangle_{4f}$ 
is obtained by multiplying the r.m.e and $\langle O^{(K)}_{2} \rangle$. 
The Stevens factors $\theta^{(K)}_{J}$ are also listed.~\cite{Stevens52} 
The absolute values of the multipolar moments are equal to $\theta^{(K)}_{J} \langle O^{(K)}_{2} \rangle$. 
Note that the value for rank 0 is nine, which is the total number of the $4f$ electrons.

This calculation shows that the rank 4 moment $(=-0.071)$ is not negligible in comparison with the rank 2 moment 
$(=-0.225)$. This can be observed in Fig.~\ref{fig10} in the difference of the charge distribution including 
up to rank 2 and rank 4 moments. An estimation of $\langle T^{(K)}_{2} \rangle_{4f}$ using the r.m.e in Table~\ref{tabl1} 
gives much larger value of $\langle T^{(4)}_{2} \rangle_{4f}$ $(=242)$ than 
$\langle T^{(2)}_{2} \rangle_{4f}$ $(=-18.5)$. However, the present experimental results, when compared with the calculated 
structure factors in Fig.~\ref{fig8}, indicate that $\langle T^{(4)}_{2} \rangle_{4f}$ is much smaller than 
$\langle T^{(2)}_{2} \rangle_{4f}$. This is inconsistent with the above simulation. 

It is possible to construct eigenfunctions which give negligibly small $\langle T^{(4)}_{2} \rangle_{4f}$, but 
such functions are not consistent with the experimental observations. 
The eigenfunctions of the ground doublet can be approximated by 
\begin{eqnarray}
\psi_a &=& c_5 |\frac{5}{2}\rangle + c_1 |\frac{1}{2}\rangle  + c_3 |-\frac{3}{2}\rangle \;,\\
\psi_b &=& c_3 |\frac{3}{2}\rangle + c_1 |-\frac{1}{2}\rangle + c_5 |-\frac{5}{2}\rangle\;,       
\end{eqnarray}
where $c_1=\cos\theta$, $c_3=\sin\theta \cos\varphi$, and $c_5=\sin\theta \sin\varphi$.~\cite{Tanaka04} 
These functions diagonalize $O^{(K)}_{2}$ for even $K$, and linear combinations of 
$(\psi_a + \psi_b)/\sqrt{2}$ and $(\psi_a - \psi_b)/\sqrt{2}$ also diagonalize $J_x$. 
$\theta=\pi/4$ and $\varphi \sim 0.22\pi$ give $\langle O^{(2)}_{2} \rangle$ and $\langle O^{(4)}_{2} \rangle$ 
around their maximum values, which are close to the values in Table~\ref{tabl1}.  
Almost identical charge distribution with that of Fig.~\ref{fig10} is obtained. 
The ratio of $\langle T^{(4)}_{2} \rangle_{4f}/\langle T^{(2)}_{2} \rangle_{4f}$ decreases with increasing $\varphi$ and 
vanishes around $\varphi=0.73\pi$. However, in this region of $\varphi$, $\langle O^{(2)}_{2} \rangle$ itself also 
becomes very small and the anisotropy of the dipolar moment within the $ab$ plane disappears; 
$\langle J_x^{\;2}\rangle$ and  $\langle J_y^{\;2}\rangle$ become to have almost the same values although 
$\langle J_z^{\;2}\rangle$ is still small. 
The charge distribution of this eigenfunction becomes cylindrical which is 
elongated along the $z$ axis. 

We consider that the dipolar moment is confined along the $x$ axis by the $O^{(2)}_{2}$ type 
quadrupolar moment because the magnetic structure in the AFM phase strongly reflects the underlying 
AFQ order. Therefore, the region around  $\varphi=0.73\pi$ is not probable. The wavefunction of the ground 
state should have a rank 4 moment that is not negligible compared with the rank 2 moment. 
The charge distribution should be like the one in Fig.~\ref{fig10}. 

However, the azimuthal-angle dependence of the RXS at the $E2$ resonance indicate that $\langle T^{(4)}_{2} \rangle_{4f}$ 
is much smaller than $\langle T^{(2)}_{2} \rangle_{4f}$.  
This disagreement could be due to the r.m.e in Table~\ref{tabl1}, which is the value calculated for the 
atomic wavefunctions within the framework of the \textit{idealized scattering length}.~\cite{Lovesey98}  
Although we consider that the rank 4 moment should be reflected in the RXS of the $E2$ resonance, 
the actual intensity should be re-examined by more realistic theoretical calculations which take into account complex 
intermediate states more properly. 
\begin{table}
\caption{Matrix element of the operator equivalent $O^{(K)}_{2}$, reduced matrix element of $T^{(K)}$ 
for the $E2$ resonance at the $L_{\text{III}}$ edge ($\bar{j}=\frac{3}{2}$), Stevens factor, and the multipolar moment 
of the calculated wavefunction in Fig.~\ref{fig10}, for rank 2 and 4. The definition of $O^{(K)}_{2}$ follows 
Ref.~\onlinecite{Shiina97}.}
\begin{ruledtabular}
\begin{tabular}{ccccc}
$K$&$\langle O^{(K)}_{2} \rangle$
&$(J||T(K:\bar{j})||J)$& $\theta^{(K)}_{J}$ & $\theta^{(K)}_{J}\langle O^{(K)}_{2} \rangle$ \\
\hline
$2$ &$35.4$& $-0.5236$ & $-6.35\times 10^{-3}$ & $-0.225$ \\
$4$ &$1196.8$& $0.2024$ & $-5.92\times 10^{-5}$ & $-0.071$ \\
\end{tabular}
\label{tabl1}
\end{ruledtabular}
\end{table}

\subsection{$d$-$f$ Coulomb and quadrupole-strain interactions}
Here, we simply assume that an electron in Dy with a wavefunction extended toward 
B and C atoms, which are negatively charged, costs high energy because of the Coulomb interaction. 
Then, the analysis of the interference effect indicates that the charge distribution of the $5d$ electron 
is determined by the $d$-$f$ Coulomb interaction, and not by the lattice distortion of B and C.  
The lattice distortion is caused by the direct interaction with the $4f$ electrons, 
not with the $5d$ electrons although they are more extended. 
This conclusion is contrary to the theoretical study of Ref.~\onlinecite{Igarashi03}, where 
it is concluded that the lattice distortion directly modulates the $5d$ states and the $E1$ resonance 
does not reflect the $4f$ quadrupolar moment. Later on, however, the same authors reconsidered the 
mechanism and reported that the $E1$ resonance can be caused by the $d$-$f$ Coulomb interaction.~\cite{Nagao03a} 

In the course of the analysis, we assumed that $F_{\text{lat}}$ and the $E1$ and $E2$ resonances 
interfere coherently; the mixing parameters $r_1$ and $r_2$ were fixed at unity, a positive real number. 
Here, it should be noted that we approximate the energy spectrum of the resonance by only two oscillators at 
$\Delta_{E1}$ and $\Delta_{E2}$, which is actually composed of many oscillators of the respective 
intermediate states. In this sense, the mixing parameter should be in general a complex number. 
Nevertheless, in view of the fact that all the experimental data, including the (0 0 2.5) reflection 
examined in the previous paper,~\cite{Matsumura04} are explained consistently by assuming the 
mixing parameters as positive real numbers, we consider that the above conclusion must be the case. 
This point also should be re-examined by a more realistic theory which properly considers the 
intermediate states.~\cite{Nagao03b}

\section{Conclusions}
We have performed RXS for the (3 0 1.5) superlattice reflection in DyB$_2$C$_2$. 
Resonant enhancement was observed below $T_Q$, indicating the AFQ order in which the principal axis of the 
quadrupolar moment is canted in accord with the site symmetry of $2/m$. 

Azimuthal-angle dependence of the intensities for the $E1$ and $E2$ resonances have been measured 
precisely. We analyzed the results using a theory which directly connects atomic tensors with the 
scattering amplitudes of RXS. The fitting results indicate that the data are well explained 
by the atomic tensors up to rank 2. The rank 4 tensor is little reflected in the data. 
However, a simulation of the wavefunction in the AFQ phase predicts a relatively large value of the rank 4 
moment in comparison with that of rank 2. There should be some reason the rank 4 moment is difficult to 
observe in the $E2$ process of RXS, which should be examined by more realistic theoretical calculations 
properly taking into account the complex intermediate states. 

Finally, an intra-atomic $d$-$f$ Coulomb interaction and a direct coupling between the $4f$ quadrupolar 
moment and the lattice have been evidenced by the analysis of the interference structure among the Thomson 
scattering from the lattice distortion, the $E2$ resonance which observe the $4f$ state, and the $E1$ resonance 
which observe the $5d$ state. The charge distributions of the $5d$ and $4f$ electrons are extended along 
the directions almost perpendicular with each other. The B and C atoms along the direction where the 
$4f$ charge distribution is extended move away from Dy, while those along the direction where the $5d$ charge 
distribution is extended move toward Dy.

\section{Acknowledgments}
The authors acknowledge Dr. S. W. Lovesey for important discussions and comments on the application of 
his theory to the analysis of our experimental results. We also thank Dr. S. Ishihara for valuable discussions. 
This study was performed under the approval of the Photon Factory Program Advisory Committee 
(Proposal No. 2001G063), and was supported by a Grant-in-Aid for Scientific Research from the Japanese Society 
for the Promotion of Science. 

\section{Appendix}
We use a formalism developed by Lovesey \textit{et al.} to analyze the experimental results of RXS. 
This formalism simply deals with the intermediate states as an atomic core hole, and then the 
resultant scattering amplitude is directly connected with an atomic tensor $\langle T^{(K)}_{q} \rangle$. 
The atomic scattering factor for a resonant process between core and valence states 
with orbital quantum numbers $\bar{l}$ and $l$, respectively, is expressed as 
\begin{eqnarray}
  Z^{\text{(E1)}}&=&  A
    \sum_{K} \sqrt{2K+1}\left\{ \begin{array}{ccc} 1 & K & 1 \\  l & \bar{l} & l \end{array}   \right\} 
    \nonumber \\
    && \times \sum_{q}\langle T^{(K)}_q \rangle X^{(K)}_{-q} (-1)^q \label{eq:ZE1}\\ 
  Z^{\text{(E2)}}&=& B
   \sum_{K}(-1)^K\sqrt{2K+1} \left\{ \begin{array}{ccc} 2 & K & 2 \\  l & \bar{l} & l \end{array}   \right\}
  \nonumber \\
  && \times \sum_{q} \langle T^{(K)}_q \rangle H^{(K)}_{-q} (-1)^q 
  \label{eq:ZE2}
\end{eqnarray}
where $A= (l || C(1) || \bar{l}) (\bar{l} || C(1) || l) \langle l | R | \bar{l} \rangle^2$ and 
$B=\frac{k k'}{6} \{ (l || C(2) || \bar{l}) \langle l | R^2 | \bar{l} \rangle \}^2$ are the constant factors 
for $E1$ and $E2$ processes, respectively. 
$ X^{(K)}_q$ and $H^{(K)}_q$ are the spherical tensors which express the state of x rays involved in the 
scattering process: 
\begin{equation}
   X^{(K)}_q = \sum_{\mu,\mu'=-1}^{1} 
         \varepsilon_{\mu}' \varepsilon_{\mu'} \langle 1 \mu 1 \mu' | K q \rangle
   \label{eq:E1X}
\end{equation}
and
\begin{eqnarray}
   h_{\mu} &=& \sum_{\nu,\nu'=-1}^{1} 
      \varepsilon_{\nu} k_{\nu'} \langle 1 \nu 1 \nu' | 2 \mu \rangle 
   \label{eq:E2smallh} \\
   H^{(K)}_q &=& \sum_{\mu,\mu'=-2}^{2} h'_{\mu'} h_{\mu} \langle 2 \mu 2 \mu' | K q \rangle \;, 
   \label{eq:E2largeH}
\end{eqnarray}
where $\bm{\varepsilon}$ ($\bm{\varepsilon}'$) and $\bm{k}$ ($\bm{k}'$) are the polarization and 
wavevectors of the incident (scattered) x ray, and they are also expressed in the form of spherical tensors of rank 1. 

The structure factor for the (3 0 1.5) reflection is calculated by $F=Z(1)-Z(2)-Z(3)+Z(4)$, where the index represents 
the four Dy ions in the unit cell. The final scattering amplitude at resonance is proportional to 
$F/(E-\Delta+i\Gamma/2)$, where $\Delta$ and $\Gamma$ are the energy and width of the resonance.


\end{document}